%% file: nature_phys.tex
\documentclass[10pt]{wlscirep}
\usepackage[resetlabels]{multibib}
\usepackage{mathtools}
\usepackage{multirow}
\usepackage{makecell}
\usepackage{booktabs}
\usepackage{array}

\usepackage{amsfonts,amssymb,stmaryrd,latexsym,amsmath,braket}
\usepackage{graphicx,subfigure}
\usepackage[export]{adjustbox}
\usepackage{times}
\usepackage{bbm}
\usepackage{bm}
\usepackage{mathrsfs}
\usepackage{slashed}
\usepackage[version=3]{mhchem}
\usepackage{braket}
\usepackage{hyperref}
\usepackage{verbatim}
\usepackage[utf8]{inputenc}

\usepackage{graphicx,subfigure}
\usepackage{times}
\usepackage{slashed}
\usepackage{braket}
\usepackage{verbatim}
\usepackage{multirow}
\usepackage[utf8]{inputenc}
\usepackage{array}
\usepackage{tensor}
\newcommand{\PreserveBackslash}[1]{\let\temp=\\#1\let\\=\temp}
\newcolumntype{C}[1]{>{\PreserveBackslash\centering}p{#1}}
\newcolumntype{R}[1]{>{\PreserveBackslash\raggedleft}p{#1}}
\newcolumntype{L}[1]{>{\PreserveBackslash\raggedright}p{#1}}

\makeatletter
\def\namedlabel#1#2{\begingroup
   \def\@currentlabel{#2}%
   \label{#1}\endgroup
}
\makeatother

\title{Multi-strangeness matter from {\em ab initio} calculations}

\author[1,*]{Hui~Tong}
\author[2,3,*]{Serdar~Elhatisari} 
\author[1,4,5,*]{Ulf-G.~Mei{\ss}ner}
\author[6,*]{Zhengxue~Ren}

\affil[1]{Helmholtz-Institut f\"{u}r Strahlen- und Kernphysik and Bethe Center for Theoretical Physics, Universit\"{a}t Bonn, D-53115 Bonn, Germany}

\affil[2]{King Fahd University of Petroleum and Minerals (KFUPM), 31261 Dhahran, Saudi Arabia}

\affil[3]{Faculty of Natural Sciences and Engineering, Gaziantep Islam Science and Technology University, Gaziantep 27010, Turkey}

\affil[4]{Institute~for~Advanced~Simulation (IAS-4),~Forschungszentrum~J\"{u}lich,~D-52425~J\"{u}lich,~Germany}

\affil[5]{Peng Huanwu Collaborative Center for Research and Education, International Institute for Interdisciplinary and Frontiers, Beihang University, Beijing 100191, China}

\affil[6]{School of Physics, Nankai University, Tianjin 300071, China}

\affil[*]{Email: tong@hiskp.uni-bonn.de, selhatisari@gmail.com, meissner@hiskp.uni-bonn.de, zxren@nankai.edu.cn}


\begin{abstract}
\input{1-abstract}
\end{abstract}

\begin{document}

\maketitle

\input{2-main}

%%%%%%%%%%%%%%%%%%%%%%%%%%%%%%%%%%%%%%%%%%%%%%
\bigskip

\noindent {\bf \large Author Contributions}
HT performed the theoretical work, led production runs, and created the figures. HT also analyzed the data with contributions by SE and ZR. The code development, testing and optimization were led by SE with contributions by HT. SE and UGM supervised the direction of the research. UGM in particular pushed towards staying within a unified framework. All authors were involved in the writing, editing, and review of this work.
\\
\\
\noindent {\bf \large Author Information}  The authors declare no competing financial interests.
\\
\\
\noindent {\bf \large Code Availability} All of the codes produced in association with this work have been stored and can be obtained upon request from the authors, subject to possible export control constraints.
\\
\\
\noindent {\bf \large Data Availability} All of the data produced in association with this work have been stored and can be obtained upon request from the authors.
\\
\\
\noindent {\bf \large Inclusion and Ethics} We have complied with community standards for authorship and all relevant recommendations with regard to inclusion and ethics.
\newpage 

%%%%%%%%%%%%%%
\section*{Methods}
\renewcommand{\thesection}{S\arabic{section}}
\renewcommand{\thefigure}{S\arabic{figure}}
\renewcommand{\theequation}{S\arabic{equation}}
\renewcommand{\thetable}{S\arabic{table}}
\setcounter{figure}{0}
\setcounter{equation}{0}
\setcounter{section}{0}
\setcounter{table}{0}

\input{3-method}

\bibliography{References}

\end{document}

%% file: 1-abstract.tex
Hypernuclei and hypernuclear matter connect nuclear structure in the strangeness sector with the astrophysics of neutron stars, where hyperons are expected to emerge at high densities and affect key astrophysical observables. 
We present the first {\em ab initio} calculations that simultaneously describe single- and double-$\Lambda$ hypernuclei from the light to medium-mass range, the equation of state for $\beta$-stable hypernuclear matter, and neutron star properties. 
Despite the formidable complexity of quantum Monte Carlo~(QMC) simulations with multiple baryonic degrees of freedom, by combining nuclear lattice effective field theory with a newly developed auxiliary-field QMC algorithm we achieve the first sign-problem free {\em ab initio} QMC simulations of hypernuclear systems containing an arbitrary number of neutrons, protons, and $\Lambda$ hyperons, including all relevant two- and three-body interactions. 
This eliminates reliance on the symmetry-energy approximation, long used to interpolate between symmetric nuclear matter and pure neutron matter.
Our unified calculations reproduce hyperon separation energies, yield a neutron star maximum mass consistent with observations, predict tidal deformabilities compatible with gravitational-wave measurements, and give a trace anomaly in line with Bayesian constraints. 
By bridging the physics of finite hypernuclei and infinite hypernuclear matter within a single {\em ab initio} framework, this work establishes a direct microscopic link between hypernuclear structure, dense matter composition, and the astrophysical properties of neutron stars.

%% file: 2-main.tex
%========================================
\section*{Introduction}
%========================================

To unravel the formation and properties of multi-strangeness matter — ranging from hypernuclei in the laboratory to hyperon-rich neutron stars in the cosmos — constitutes the central goal of strangeness nuclear physics~\cite{Rezzolla:2018jee,Schaffner-Bielich:2020psc}. 
By incorporating strangeness, these systems extend nuclear structure and equation of state (EoS) studies into a new regime, connecting the familiar up–down sector of Quantum Chromodynamics~(QCD) with its strange quark sector~\cite{Povh:1978mx,Gal:2016boi,Lattimer2016,Oertel:2016bki,Baym:2017whm,Tolos:2020aln,Burgio:2021vgk,Sedrakian:2022ata,MUSES:2023hyz,Haidenbauer:2025zrr}. 
While remarkable progress has been achieved in hypernuclear physics, a precise understanding of interactions involving hyperons ($Y$) still lags far behind that of two- and three-nucleon ($NN$, $3N$) interactions. 
This deficiency stems from the extreme scarcity of low-energy $YN$ and hyperon-hyperon ($YY$) scattering data, together with hypernuclei measurements restricted to only a few dozen single-$\Lambda$ and a handful of double-$\Lambda$ systems, predominantly in the light-mass region~\cite{Haidenbauer:2015zqb,Gal:2016boi,J-PARCE40:2021bgw,Eckert2023,Miwa:2025adw,Haidenbauer:2025zrr}. 
This narrow experimental foundation leaves the fundamental two- and three-body hyperonic forces essentially unconstrained. 
This uncertainty hinders the establishment of a unified framework that bridges the physics of finite hypernuclei — from light to medium–mass systems — to the properties of infinite, dense baryonic matter.

Early microscopic studies focused on the light hypernuclei, beginning with Faddeev calculations of the hypertriton~\cite{Miyagawa:1993rd,Miyagawa:1995sf}, later extended to $\prescript{4}{\Lambda}{}$H, $\prescript{4}{\Lambda}{}$He, and $\prescript{4}{\Lambda\Lambda}{}$H using the Faddeev–Yakubovsky formalism~\cite{Nogga:2001ef,Filikhin:2002wp} and variational methods~\cite{Hiyama:2001zt,Nemura:2002fu}. 
Within cluster-model approaches, the range could be pushed to $A=10$-$13$ $\Lambda$ hypernuclei~\cite{Hiyama:2000jd,Hiyama:2012sq}, but only by employing simplified phenomenological effective potentials. 
In the past decade, theoretical and computational advances have substantially expanded the scope of microscopic hypernuclear studies~\cite{Haidenbauer:2025zrr}. 
Modern \textit{ab initio} calculations, such as the no-core shell model~(NCSM), now permit calculations of binding energies beyond the $s$-shell up to $\prescript{13}{\Lambda}{}$C~\cite{Wirth:2014apa,Wirth:2016iwn,Wirth:2017bpw,Le:2020zdu,Knoll:2023mqk}, and a notable recent advance is that hypernuclei up to $A=7$ can be quantitatively described once chiral $\Lambda NN$ and $\Sigma NN$ three-body forces are incorporated consistently ~\cite{Le:2024rkd}. 
The stochastic variational method~(SVM) is also a widely used method for studying light $\Lambda$ hypernuclei~\cite{Nemura:1999qp,Contessi:2019csf,Schafer:2020rba,Schafer:2021qbt,Schafer:2022une}, and has recently been employed within pionless effective field theory~(EFT) to successfully address the long-standing overbinding problem of $\prescript{5}{\Lambda}{}$He~\cite{Contessi:2018qnz}. 
Quantum Monte Carlo (QMC) methods, including variational Monte Carlo (VMC)~\cite{Usmani:2004ts,Usmani:2006jd} and auxiliary-field diffusion Monte Carlo (AFDMC)~\cite{Lonardoni:2012rn,Lonardoni:2013rm,Lonardoni:2013gta}, have been employed to study light- and medium-mass $\Lambda$ hypernuclei, albeit relying on simplified phenomenological interactions. 
More recently, the properties of light hypernuclei have been investigated using a VMC approach with neural network quantum states (VMC-QNS) in a pionless EFT framework, yielding good agreement with separation energies up to $\prescript{16}{\Lambda}{}$O~\cite{DiDonna:2025oqf,Zhang:2025okd}. 
Furthermore,  nuclear lattice effective field theory (NLEFT)~\cite{Elhatisari:2022zrb,Hildenbrand:2024ypw}, a powerful quantum many-body approach that combines EFT with lattice Monte Carlo techniques, using high-fidelity $2N$ and $3N$ interactions at next-to-next-to-next-to-leading-order (N${}^3$LO) together with enhanced leading-order (LO) $\Lambda N$ interaction and $\Lambda NN$ interactions at next-to-next-to-leading-order (N${}^2$LO) has also been applied to hypernuclei up to $\prescript{16}{\Lambda}{}$O. 

In contrast to finite hypernuclei, where strangeness is introduced by adding one or two $\Lambda$ hyperons to the nuclear core, the interior of neutron stars may undergo a much more dramatic transformation at supra-nuclear saturation densities~\cite{Tolos:2020aln,Burgio:2021vgk,Sedrakian:2022ata,MUSES:2023hyz}. 
As the density increases above twice nuclear saturation density, $2\rho_0$ with $\rho_0 = 0.16\,$fm$^{-3}$, it becomes energetically favorable for neutrons to convert into hyperons through weak processes~\cite{Tolos:2020aln,Burgio:2021vgk}. 
This onset of hyperons provides additional degrees of freedom and generally softens the EoS of dense matter, lowering the pressure for a given energy density. 
While such a mechanism is a natural consequence of baryonic many-body theory, it creates a striking tension with astrophysical observations: most microscopic and phenomenological approaches that allow for hyperons predict a strong softening of the EoS, resulting in maximum neutron star masses that fall below the observed $\sim2M_\odot$ pulsars. 
This inconsistency, known as the {\em hyperon puzzle}, has become one of the central open problems at the interface of nuclear physics and astrophysics~\cite{Schaffner-Bielich:2008zws,Takatsuka:2008zz,Djapo:2008au,Vidana:2010ip,Schulze:2011zza,Weissenborn:2011ut,Massot:2012pf,Yamamoto:2013ada,Yamamoto:2014jga,Astashenok:2014pua,Lonardoni:2014bwa,Maslov:2015msa,Chatterjee:2015pua,Masuda:2015kha,Haidenbauer:2016vfq,Fortin:2017cvt,Logoteta:2019utx,Gerstung:2020ktv,Schaffner-Bielich:2020psc,Li:2020dst,Friedman:2022bpw,CompOSECoreTeam:2022ddl,Ye:2024meg,Bauswein:2025dfg,Ferreira:2025dat,Vidana:2024ngv}. 
Among the microscopic theoretical calculations, Brueckner–Hartree–Fock (BHF) theory has been widely employed to calculate neutron star properties using hyperonic interactions derived from meson-exchange models, chiral EFT, or even Lattice QCD~\cite{Hu:2014wja,HALQCD:2019wsz,Logoteta:2019utx,Vidana:2024ngv,Zheng:2025sol}. 
The inclusion of genuine $3N$ forces in the BHF theory remains a challenging task, as it would require solving the three-body Bethe–Faddeev equation in the nuclear medium~\cite{Vidana:2024ngv}. 
The BHF theory also rests on the two-hole-line expansion and is sensitive to the choice of the single-particle potential, thereby introducing uncertainties into many-body calculations~\cite{Wang:2022xlr}. 
This uncertainty can be substantially reduced by including contributions from the three-hole-line expansions~\cite{Song:1998zz,Lu:2017nbi}.
In Ref.~\cite{Logoteta:2019utx},  hypernuclear matter (HNM) was calculated within the BHF theory, but a perturbative many-body approach has been used to compute the $\Lambda$ separation energies for $^{41}_\Lambda$Ca, $^{91}_\Lambda$Zr, and $^{209}_\Lambda$Pb. For the hypernuclei $^{91}_\Lambda$Zr and $^{209}_\Lambda$Pb, including the $NN\Lambda$ interaction significantly improves the agreement between theory and experiment. 
In contrast, for $^{41}_\Lambda$Ca the two $NN\Lambda$ models introduce excessive repulsion and fail to reproduce the experimental $\Lambda$ separation energy, a trend also seen in lighter hypernuclei. 
While consistent $\chi$EFT-based $NN$ and $3N$ interactions are used, the available $\chi$EFT $N\Lambda$ forces were not yet implemented in their framework, leading them to adopt the Nijmegen meson-exchange interaction, although for consistency one would expect hyperonic interactions derived within the same framework. 
In addition, the groundbreaking AFDMC calculations have been carried out for hyperneutron matter consisting of a limited number of neutrons and $\Lambda$ hyperons, employing simplified phenomenological interactions in which the $NN\Lambda$ force was tuned to reproduce $\Lambda$ separation energies of selected hypernuclei~\cite{Lonardoni:2013rm,Lonardoni:2013gta,Lonardoni:2014bwa}. 
The authors concluded that the appearance of hyperons in neutron star interiors could not be firmly established and was not necessarily in conflict with astrophysical observations, while emphasizing that hyperonic three-body forces require further investigation within frameworks such as chiral perturbation theory~\cite{Kaiser:2004fe,Haidenbauer:2013oca}.

Despite such progress, a critical limitation remains in all current {\em ab initio} Monte Carlo simulations. The neglect of protons precludes a consistent treatment of $\beta$-stable HNM~\cite{Lonardoni:2014bwa,Tong:2024jvs,Tong:2025sui}. This omission arises from the formidable complexity of handling multiple baryonic degrees of freedom in QMC simulations. 
Yet protons play a crucial role in maintaining chemical equilibrium and charge neutrality in neutron star matter, making their inclusion essential for a realistic description.

Given the scarcity of experimental constraints and the shortcomings of existing theoretical methods, there has been a long-standing anticipation for an {\em ab initio} approach that can simultaneously describe hypernuclei across the nuclear chart and the dense HNM in neutron stars. 
In this study, we present the first {\em ab initio} calculations that unify the description of finite hypernuclei, $\beta$-stable HNM, and neutron star properties. 
Building on NLEFT~\cite{Lee:2008fa,Lahde:2019npb}, we employ a newly developed auxiliary-field quantum Monte Carlo (AFQMC) algorithm~\cite{Tong:2024jvs,Tong:2025sui,Hildenbrand:2024ypw} that enables sign-problem free simulations of systems with an arbitrary number of neutrons, protons, and $\Lambda$ hyperons, thereby eliminating the need for the symmetry-energy approximation traditionally used to interpolate between symmetric nuclear matter~(SNM) and pure neutron matter~(PNM). 
Our framework incorporates all relevant two- and three-body hyperonic forces—$N\Lambda$, $\Lambda\Lambda$, $NN\Lambda$, and $N\Lambda\Lambda$—with hyperonic couplings constrained by scattering data and $\Lambda$ separation energies of single- and double-$\Lambda$ light hypernuclei. 
With these inputs, we predict $\Lambda$ separation energies across a wide mass range, determine the EoS of $\beta$-stable HNM up to several times nuclear saturation density, and compute neutron star observables including the mass–radius relation and tidal deformabilities. 
In doing so, we establish a direct microscopic link between hypernuclear structure, dense matter composition, and astrophysical observations, offering a controlled resolution of the long-standing {\em hyperon puzzle} within a single {\em ab initio} framework.

%========================================
\section*{Results and discussion}
%========================================

We present results obtained from nuclear lattice simulations using the following Hamiltonian,
\begin{equation}
  H = H_{\rm free} + V_{NN} + V_{N\Lambda} + V_{\Lambda\Lambda} + V_{NNN} + V_{NN\Lambda} + V_{N\Lambda\Lambda},
  \label{eq:Hamiltonian}
\end{equation}
where $H_{\rm free}$ is the kinetic energy term for nucleons and $\Lambda$ hyperons. 
The $NN$ and $NNN$ interactions employed here were tuned in our previous work~\cite{Tong:2024jvs,Tong:2025sui} to experimental scattering phase shifts and the saturation properties of symmetric nuclear matter~(SNM). 
These interactions also accurately describe ground-state energies of finite nuclei~\cite{Tong:2024jvs,Tong:2025sui}. 
The forms of the two- and three-body interactions are detailed in the Methods section. 
This Hamiltonian provides the microscopic starting point for our unified description of both finite hypernuclei and dense hypernuclear matter~(HNM).

\begin{figure}[htbp] 
  \centering\includegraphics[width=1.0\textwidth,valign=t]{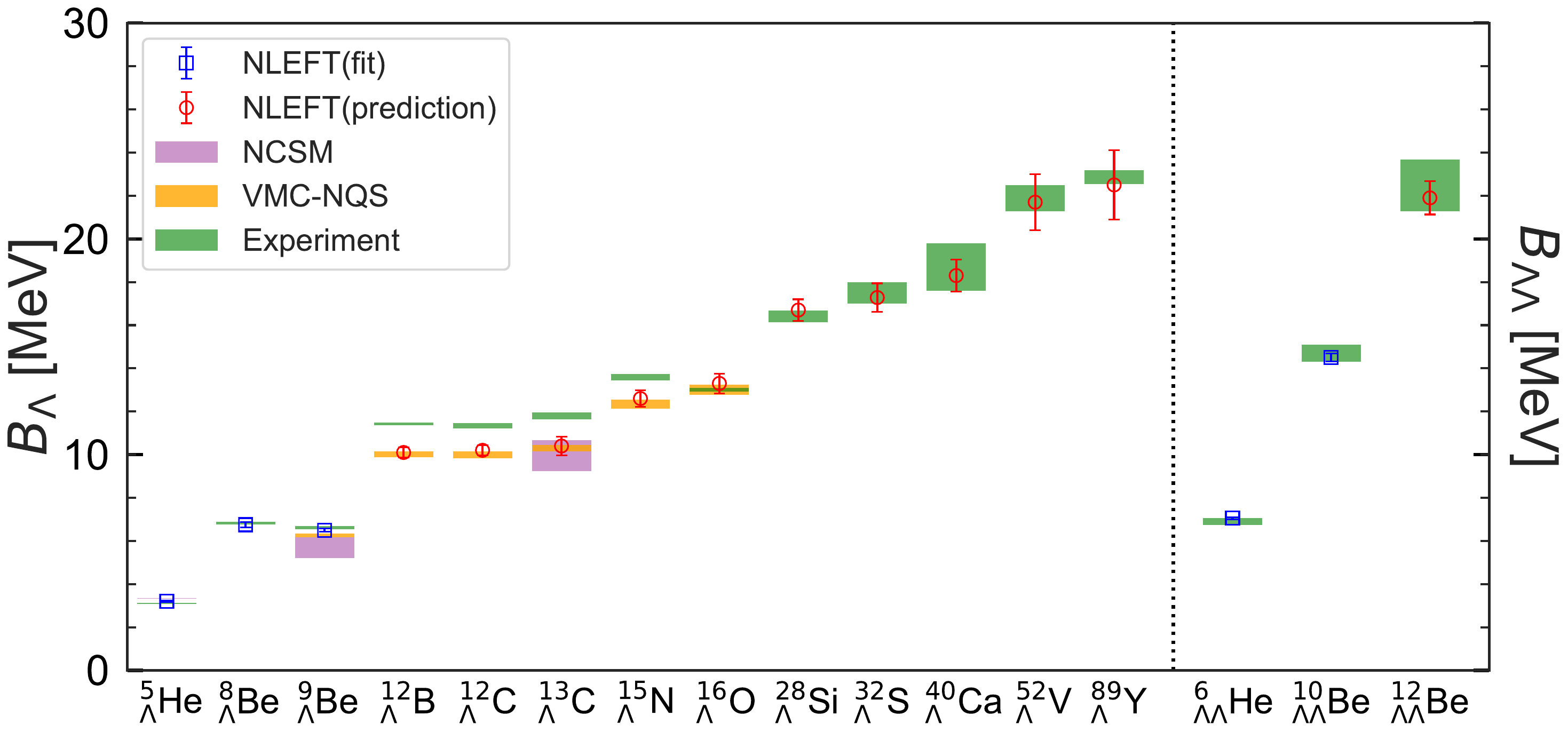}
  \caption{ {\bf Single-$\Lambda$ separation energies and double-$\Lambda$ separation energies for hypernuclei.} Hypernuclei used to fit the $NN\Lambda$ and $N\Lambda\Lambda$ forces are depicted by blue squares. Predictions are shown as red circles for single-$\Lambda$ hypernuclei up to $_\Lambda^{89}$Y and double-$\Lambda$ hypernuclei up to $_{\Lambda\Lambda}^{12}$Be. Error bars represent the combined statistical and systematic uncertainties. NLEFT results are compared with experimental data from the Hypernuclear Database (green)~\cite{Eckert2023}, and with recent {\em ab initio} predictions from the variational Monte Carlo method based on neural network quantum states (VMC-NQS, orange)~\cite{DiDonna:2025oqf} and the no-core shell model (NCSM, purple)~\cite{Knoll:2023mqk}. 
  \label{fig1}}
\end{figure}

Figure~\ref{fig1} shows the single-$\Lambda$ separation energies $B_\Lambda$ and double-$\Lambda$ separation energies $B_{\Lambda\Lambda}$ across a wide range of hypernuclei. 
The experimental data of hypernuclei used in our NLEFT calculations to constrain the $NN\Lambda$ and $N\Lambda\Lambda$ forces (blue squares) are reproduced with high accuracy, while our predictions (red circles) extend up to $_\Lambda^{89}$Y for single-$\Lambda$ systems and up to $_{\Lambda\Lambda}^{12}$Be for double-$\Lambda$ systems. 
The theoretical uncertainties include both the statistical and systematic errors.
This provides, for the first time, a coherent description of both light and medium-mass hypernuclei within an {\em ab initio} QMC framework. 
To assess the robustness of our results, the NLEFT calculations are benchmarked against two recent {\em ab initio} approaches and the available experimental data (green bands). 
Ref.~\cite{Knoll:2023mqk} employs the importance-truncated NCSM (purple) with chiral $2N+3N$ interactions at N$^3$LO and optimized $N\Lambda$ interaction at LO, but without explicit $NN\Lambda$ three-body forces. 
On the other hand, Ref.~\cite{DiDonna:2025oqf} utilizes the VMC method with neural network quantum states (VMC–NQS, orange), based on an improved leading-order pionless EFT expansion. 
Currently, the VMC–NQS method is limited to light systems up to $\prescript{16}{\Lambda}{}$O, where the separation energy of $\prescript{16}{\Lambda}{}$O is also used to constrain the $NN\Lambda$ three-body force. 
Within the light hypernuclei up to $\prescript{16}{\Lambda}{}$O, our results are in very good agreement with both NCSM and VMC–NQS within uncertainties. 
In particular, the VMC–NQS calculations and NLEFT yield quantitatively similar values, though both approaches tend to underestimate the $\Lambda$ separation energies of $\prescript{12}{\Lambda}{}$B, $\prescript{12}{\Lambda}{}$C, $\prescript{13}{\Lambda}{}$C, and $\prescript{15}{\Lambda}{}$N, a discrepancy that is expected to be resolved in future work by including higher-order interactions. 
Interestingly, unlike VMC–NQS, which explicitly uses $\prescript{16}{\Lambda}{}$O to constrain the $NN\Lambda$ three-body force, our results achieve excellent agreement with experimental data without fitting to this system, indicating that the essential physics of the $NN\Lambda$ three-body force is already captured in our formulation. 
This not only strengthens the predictive power of NLEFT, but also opens a promising avenue for extrapolations towards heavier hypernuclei. 
Consequently, we are able to present predictions up to $\prescript{89}{\Lambda}{}$Y, where experimental information is currently very scarce. 
Remarkably, even in this region our results remain consistent with the limited data within uncertainties. 

Double-$\Lambda$ hypernuclei provide the most direct systems for probing the strangeness $S=-2$ baryon–baryon interaction, yet experimental information on them is even scarcer than that for single-$\Lambda$ hypernuclei. 
Recent experimental advances, including precise hyperon–nucleon scattering measurements and the observation of double-strangeness hypernuclei in emulsion experiments at KEK and J-PARC, now offer valuable constraints on $\Lambda\Lambda$ forces~\cite{Miwa:2025adw}. 
Here we therefore provide predictions only for $\prescript{12}{\Lambda\Lambda}{}$Be, and our result shows excellent agreement with the available data within uncertainties. 
These demonstrate that the fitted $NN\Lambda$ and $N\Lambda\Lambda$ forces capture the key features of hypernuclei binding across the nuclear chart. 
These results also provide valuable guidance for future hypernuclei experiments and establish robust constraints for constructing the EoS of hypernuclear matter at high densities—thereby directly addressing the long-standing {\em hyperon puzzle} in neutron star physics.

Next, using our state-of-the-art nuclear lattice simulations and forces described above we predict the EoSs of $\beta$-stable nuclear matter~(NM) and HNM. 
For NM, our framework allows calculations with arbitrary number of protons without relying on external tools such as Gaussian process emulators~\cite{Keller:2022crb,Keller:2024snt}, and it eliminates the dependence on the commonly used symmetry-energy approximation, i.e., interpolations between SNM and pure neutron matter~(PNM). 
In our calculations, the energy per nucleon for SNM at the saturation point is $-16.90(0.02)(0.25)$~MeV, where the first parenthesis represents the statistical error and the second denotes the theoretical uncertainty quantifying the smearing effects on the SU(4) symmetric three-nucleon forces, with details provided in the Methods section. 
We predict the compression modulus $K_\infty = 229.0(3.6)$~MeV, in good agreement with the empirical value of $K_\infty = 240(20)$~MeV \cite{Garg:2018uam}. 
The accurate symmetry energy $E_{\rm sym}$ at the saturation density is $33.8(6)\,$~MeV, which agrees with the ranges of (30-35)~MeV~based on chiral EFT interactions~\cite{Hebeler:2009iv,Tews:2012fj,Holt:2016pjb,Drischler:2017wtt,Jiang:2020the,Keller:2022crb,Lim:2023dbk}.

\begin{figure}[htbp]
  \centering\includegraphics[width=0.5\textwidth,valign=t]{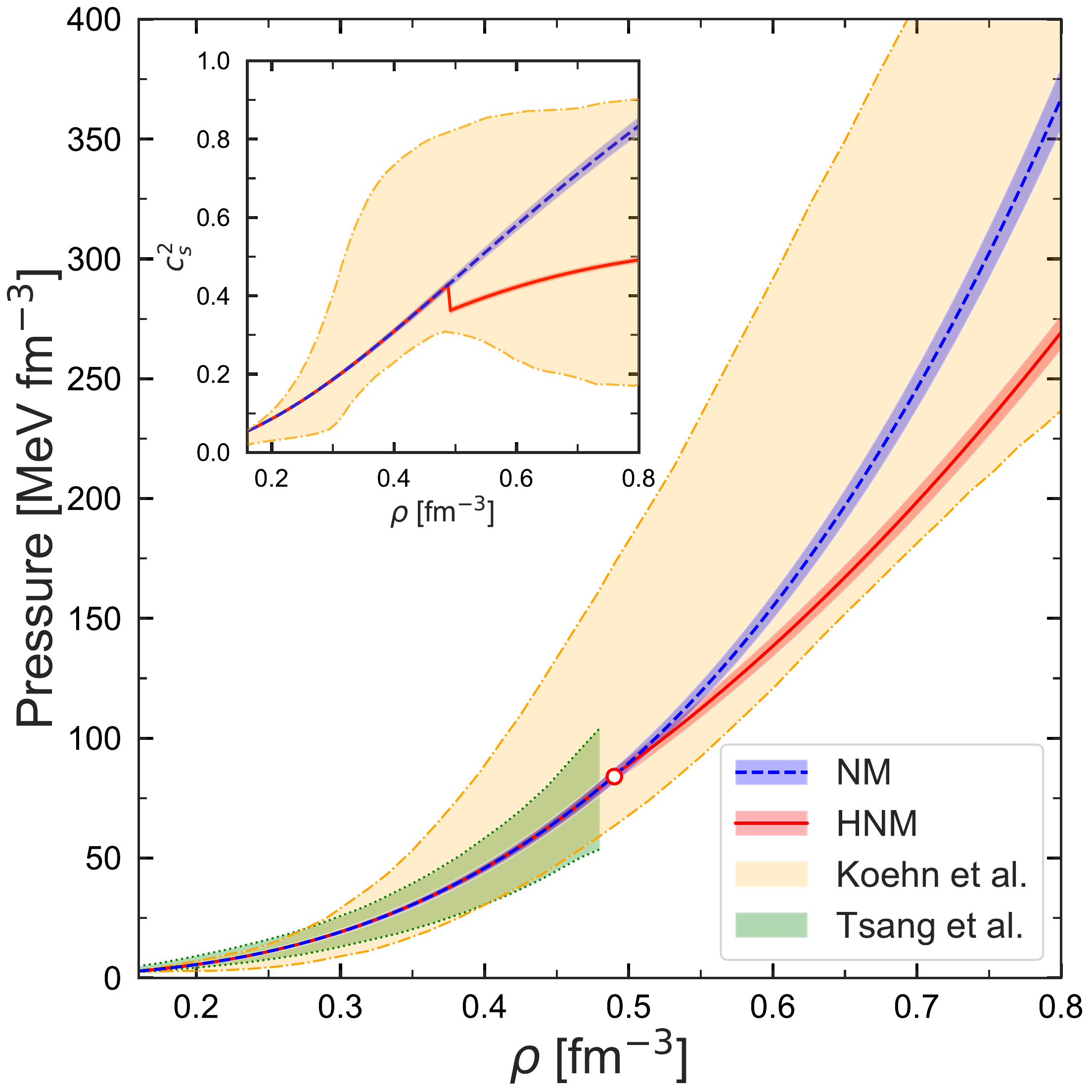}
  \caption{ {\bf EoSs of $\beta$-stable nuclear matter (NM) and hypernuclear matter~(HNM).} The blue dashed line denotes the $\beta$-stable NM, obtained from the $NN$ and $NNN$ interactions. The red solid line represents the HNM EoS including hyperons interacting via two-body ($N\Lambda$, $\Lambda\Lambda$) and three-body forces ($NN\Lambda$, $N\Lambda\Lambda$). Open circle indicates the $\Lambda$ threshold densities, $\rho_\Lambda^{\rm th}$. The inset displays the corresponding speed of sound for NM and HNM. The orange shaded regions show the final posterior estimate for the neutron star EoS constrained by $\chi$EFT, pQCD, and astrophysical observations~\cite{Koehn:2024set}, while the green shaded regions represent the posterior estimate constrained by astronomical observations together with nuclear experimental data~\cite{Tsang:2023vhh}.
  \label{fig2}}
\end{figure}

Now we turn to the discussion of the HNM system consisting of neutrons, protons, and $\Lambda$ hyperons. In this work, we have, for the first time, performed {\em ab initio} MC simulations of fully mixed systems with arbitrary number of neutrons, protons, and $\Lambda$ hyperons. 
Carrying out such simulations with multiple fermionic degrees of freedom is an exceptionally demanding task; nonetheless, this has become feasible through our novel AFQMC formulation, which transcends the limitations of earlier AFDMC studies~\cite{Lonardoni:2014bwa} based on parameterized interpolations between simpler systems. 
We denote the densities of neutrons, protons, and $\Lambda$ hyperons with $\rho_n$, $\rho_p$, and $\rho_\Lambda$, respectively, and total baryon density is $\rho=\rho_n+\rho_p+\rho_\Lambda$. 
The corresponding particle fractions in this system are $x_p=\rho_p/\rho$ and $x_\Lambda=\rho_\Lambda/\rho$. 
At each baryon density, $\beta$-equilibrium requires the particle fractions $x_i=\rho_i/\rho$ to satisfy the conditions
\begin{equation}\label{equ-chemicalequilibrium}
  \mu_n = \mu_\Lambda, \quad \mu_n - \mu_p = \mu_e, \quad \mu_e = \mu_\mu,
\end{equation}
together with charge neutrality, $\rho_p = \rho_e + \rho_\mu$. 
$\mu_i$ is the chemical potential of particle species $i$. 
Here, the corresponding chemical potentials $\mu_n$, $\mu_p$, and $\mu_\Lambda$ are obtained from the thermodynamic relation
\begin{equation}\label{equ_chemical}
  \mu_n(\rho,x_p,x_\Lambda)=\frac{\partial \varepsilon_{\scriptscriptstyle \rm HNM}}{\partial \rho_n}, ~~\mu_p(\rho,x_p,x_\Lambda)=\frac{\partial \varepsilon_{\scriptscriptstyle \rm HNM}}{\partial \rho_p}, ~~\mu_\Lambda(\rho,x_p,x_\Lambda)=\frac{\partial \varepsilon_{\scriptscriptstyle \rm HNM}}{\partial \rho_\Lambda},
\end{equation}
where $\varepsilon$ is the energy density of the system. 
This indicates that an accurate determination of the chemical potentials necessitates computing the energy density for various densities and different number of neutrons, protons, and $\Lambda$ hyperons. Fig.~\ref{figS1} shows the energy per baryon of HNM as a function of the proton number $N_p$ and the $\Lambda$ hyperon number $N_\Lambda$ at fixed density $\rho=0.6$~fm$^{-3}$.

In contrast to previous works that considered only neutrons and $\Lambda$ hyperons in the composition of neutron star matter~\cite{Lonardoni:2014bwa,Tong:2024jvs,Tong:2025sui}, this study includes the contribution of protons, whose fraction in neutron star interiors may exceed 10\%  reported previously~\cite{Tong:2022yml,Bombaci:2018ksa,Logoteta:2019utx}. 
Determining the equilibrium composition of $\beta$-stable HNM within an {\em ab initio} MC calculation requires mapping the full three-dimensional space of neutrons, protons, and $\Lambda$ hyperons -- a huge challenge that our AFQMC approach makes tractable. 
Figure~\ref{figS2} presents the resulting particle fractions as functions of density. 
As the density increases, the proton fraction rises to above $\sim 10\%$ before gradually declining at higher densities, a behavior that is in full qualitative agreement with earlier microscopic calculations based on BHF theory~\cite{Logoteta:2019utx,Vidana:2024ngv}.

\begin{figure}[htbp]
  \includegraphics[height=8.4 cm,valign=t]{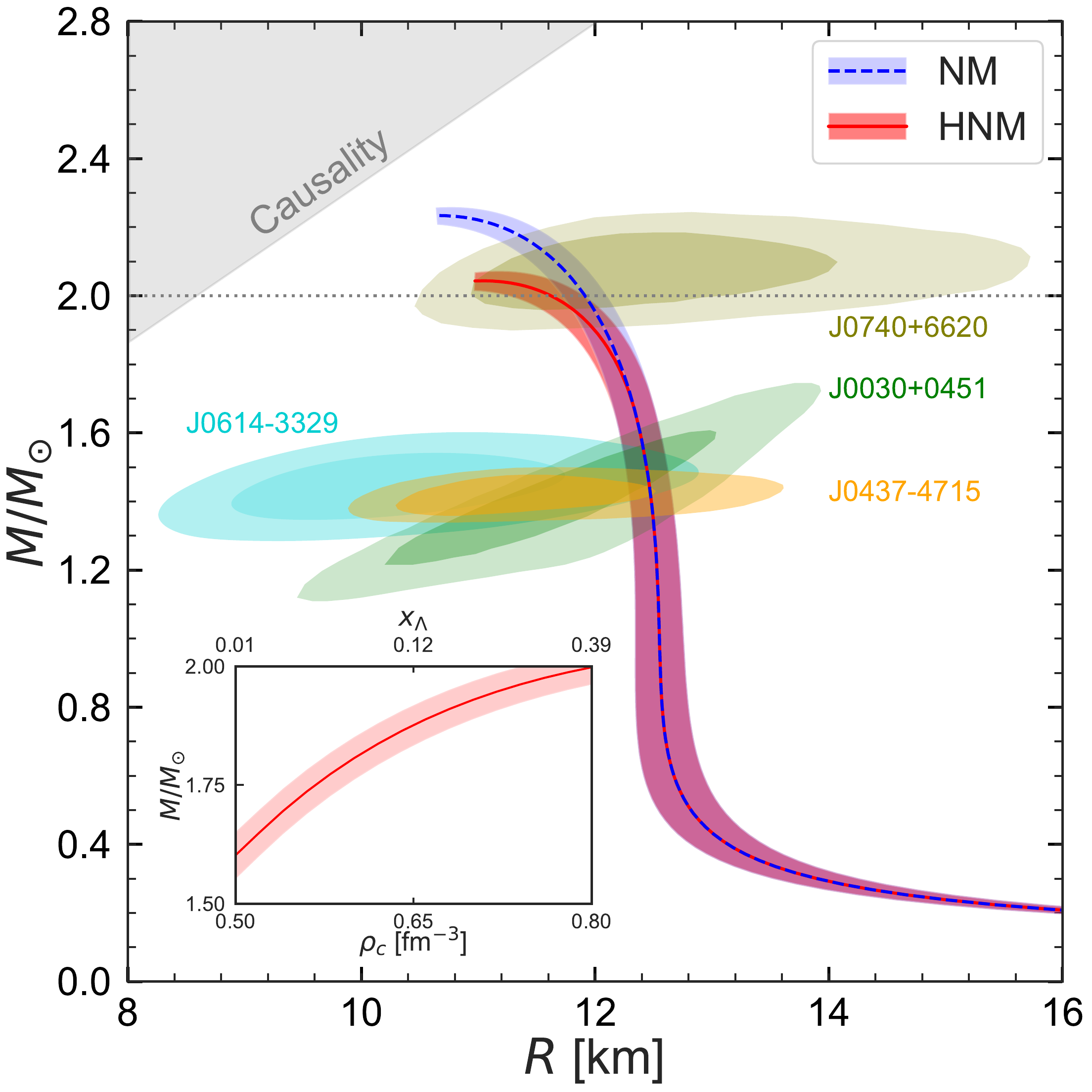}~~~~
  \includegraphics[height=8.4 cm,valign=t]{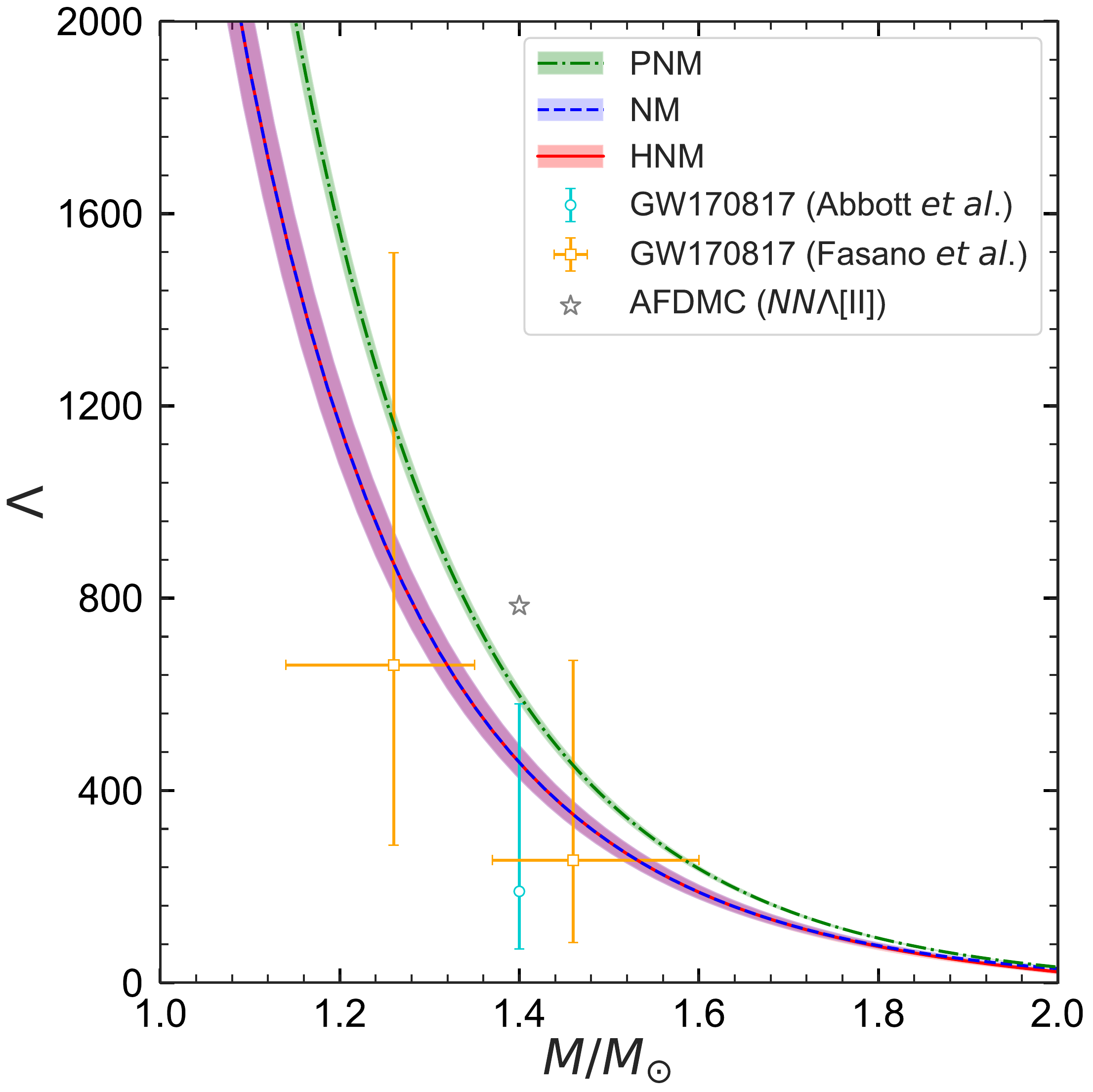}
  \caption{{\bf Left Panel: Neutron star mass-radius relation.} The legend is the same as of Fig.~\ref{fig2}. The gray horizontal dotted line marks 2$M_\odot$. The inner and outer shaded contours indicate the mass-radius constraints from NICER’s analysis of PSR~J0030+0451~\cite{Vinciguerra2024APJ}, PSR~J0740+6620~\cite{Salmi2024APJ}, PSR~J0437-4715~\cite{Choudhury2024APJ}, and PSR~J0614-3329~\cite{Mauviard:2025dmd}. The inset shows the neutron star mass as a function of central density, together with the corresponding $\Lambda$-hyperon particle fractions. {\bf Right Panel: Neutron star tidal deformability $\Lambda$ as a function of mass.} $\Lambda(M)$ is compared with the masses and tidal deformabilities inferred in Ref.~\cite{Fasano:2019zwm} for the two neutron stars in the merger event GW170817 (open squares) as well as $\Lambda(1.4M_\odot)$ extracted from GW170817~\cite{LIGOScientific:2018cki}~(open circle). The green dot-dashed line represents the pure neutron matter~(PNM) from the previous NLEFT calculations~\cite{Tong:2025sui}. The gray star denotes the result from auxiliary-field diffusion Monte Carlo (AFDMC) using their parameterization~(II) of the $NN\Lambda$ force.~\cite{Lonardoni:2014bwa}.}
  \label{fig3}
  \end{figure}

In Fig.~\ref{fig2}, the EoSs for $\beta$-stable NM and HNM are displayed. 
The threshold density for the onset of $\Lambda$ hyperons is found to be $\rho_\Lambda^{\rm th} = 0.490(0.002)(0.005)$~fm$^{-3}$. 
This value is consistent with several phenomenological schemes \cite{Weber:1989uq,Balberg:1997yw,Tolos:2016hhl} and microscopic models \cite{Baldo:1998hd,Djapo:2008au,Schulze:2011zza}, which generally predict hyperon formation in the inner core of neutron stars at $\rho \approx (2$–$3)\rho_0$. 
As anticipated, the inclusion of $\Lambda$ hyperons leads to a softening of the EoS. 
This effect is directly tied to the absence of Pauli blocking for hyperons, which allows the system to lower its Fermi pressure at high densities. 
The conversion of neutrons into $\Lambda$’s redistributes baryonic density into new degrees of freedom, reducing the pressure support against gravity.
The squared speed of sound, $c_s^2$, displayed in the inset of Fig.~\ref{fig2}, also reflects this transition: while purely nucleonic matter exhibits a monotonic increase in $c_s^2$ with density, the onset of hyperons induces non-monotonic behavior and a sharp reduction of $c_s^2$ at $\rho_\Lambda^{\rm th}$. 
This drop further signals a softening of the EoS and identifies the density regime where hyperonic degrees of freedom become dynamically relevant. 
Importantly, the causality bound ($c_s^2 < 1$) is satisfied in both NM and HNM. 
For comparison, the final posterior estimate for the neutron star EoS and $c_s^2$, constrained by chiral EFT ($\chi$EFT), perturbative QCD (pQCD), and astrophysical observations, are shown as the orange shaded regions~\cite{Koehn:2024set}. 
The green shaded regions are obtained by performing Bayesian analyses of 12 nuclear experimental constraints with three complementary constraints from astronomical observation, and an EoS was constructed from metamodelling in the form of a Taylor expansion around the saturation density up to fourth order~\cite{Tsang:2023vhh}. 
Although the inclusion of $\Lambda$ hyperons inevitably softens the EoS, we find, quite remarkably, that our HNM results remain fully consistent with these posterior probability distributions. 
This agreement is particularly interesting since the latter omit hyperonic degrees of freedom in the microscopic composition of the EoS, whereas our calculation explicitly includes them, underscoring the robustness of the microscopic consistency. 
In addition to the sound velocity, the normalized trace anomaly $\Delta=1/3-P/\varepsilon$ serves as a key measure of conformality in dense matter~\cite{Fujimoto:2022ohj,Annala:2023cwx,Cai:2024oom}. 
Figure~\ref{figS3} shows that $\Delta$ is positive at low densities, decreases with increasing density, and approaches zero at supranuclear densities, indicating a gradual trend toward conformal behavior at supra-nuclear densities. 
The inclusion of $\Lambda$ hyperons shifts $\Delta$ upward relative to NM, reflecting the softening of the EoS due to additional degrees of freedom. 
Our results remain within the credible intervals from Bayesian analyses constrained by $\chi$EFT, pQCD, and astrophysical observations~\cite{Annala:2023cwx}, highlighting both the robustness of our {\em ab initio} approach and the quantitative role of hyperons. The consistency with astrophysical posteriors arises from the fact that, in our framework, the three-body hyperonic interactions are systematically constrained by single- and double-$\Lambda$ separation energies of hypernuclei, cf. Fig.~\ref{fig1}. These constraints ensure that the additional hyperonic degrees of freedom soften the EoS in a controlled manner. At the same time, the repulsive components of the three-body forces counterbalance this effect. In this way, our approach provides a solid microscopic foundation for reliable predictions of neutron star properties that incorporate hyperonic matter.

Figure~\ref{fig3} displays the mass–radius relation (left panel) and tidal deformabilities (right panel) of neutron stars obtained from our NLEFT calculations of $\beta$-stable NM and HNM. 
These relations for NM and HNM are obtained from the EoSs of Fig.~\ref{fig2}. 
The maximum mass for NM and HNM are 2.23(0.01)(0.02)~$M_\odot$ and 2.04(0.01)(0.02)~$M_\odot$. 
These are in agreement with the calculation performed in Refs.~\cite{Lonardoni:2014bwa,Logoteta:2019utx,Han:2022rug}. 
A first striking feature is that, although the inclusion of $\Lambda$ hyperons significantly reduces the stiffness of the EoS and lowers the maximum mass compared with the purely nucleonic case, the predicted curves remain fully compatible with the massive neutron star observations~\cite{Demorest:2010bx,Fonseca:2016tux,NANOGrav:2017wvv,Salmi2024APJ}. 
This demonstrates that the $NN\Lambda$ and $N\Lambda\Lambda$ three-body forces are sufficient to counteract the hyperon-induced softening, thereby resolving the apparent tension between hyperon onset and the existence of massive neutron stars. 
The inset of the left panel shows the evolution of neutron star mass as a function of density $\rho_c$ at the stellar center and its corresponding $\Lambda$ particle fraction. 
We find that the $\Lambda$ fraction rises steadily with increasing mass, reaching as high as 39\% in the core of a $2M_\odot$ star. 
This result stands in contrast to previous AFDMC studies~\cite{Lonardoni:2014bwa}, which concluded that in order to support $2M_\odot$ neutron stars, their interiors must remain free of $\Lambda$ hyperons. 
Our findings therefore indicate that massive neutron stars can indeed host substantial hyperonic components in their interiors, with important consequences for their thermal and rotational evolution. 
The predicted mass–radius relation also agrees well with recent NICER measurements of PSR~J0030+0451, PSR~J0740+6620, PSR~J0437–4715, and PSR~J0614-3329, as indicated by the observational contours in Fig.~\ref{fig3}. 
This consistency highlights the robustness of our EoS and provides an {\em ab initio} confirmation that hyperonic degrees of freedom can be reconciled with current constraints on neutron star properties.

Next, we consider the tidal deformability (right panel), which is another important constraint of neutron star properties in the multimessenger era~\cite{LIGOScientific:2017vwq}. 
As a comparison, the results of PNM obtained by NLEFT are also shown. 
It can be observed that the PNM EoS is overly stiff and yields $\Lambda_{1.4M_\odot}$ values above the ranges inferred from GW170817.
By contrast, once protons are consistently included in $\beta$-equilibrium, the EoS softens and the predicted tidal deformabilities fall in line with the observational constraints~\cite{LIGOScientific:2018cki,Fasano:2019zwm}. 
The gray star denotes the result obtained from AFDMC calculations using their second parametrization of the $NN\Lambda$ three-body force~\cite{Lonardoni:2014bwa}, tuned to reproduce the separation energy up to $A=91$ hypernuclei; however, since only neutrons and $\Lambda$ hyperons were included, the resulting EoS is also overly stiff and clearly incompatible with astrophysical observations. 
These underscore the importance of going beyond the simplified picture of PNM in QMC calculations: $\beta$-stable matter with protons and hyperons not only matches astrophysical data more closely but also provides a more realistic description of neutron star interiors.

In summary, we have presented the first {\em ab initio} calculations that achieve a unified description of single- and double-$\Lambda$ hypernuclei from light to medium-mass systems, the EoS of $\beta$-stable hypernuclear matter, and neutron star properties. 
By combining NLEFT with a newly developed AFQMC algorithm, we have carried out the first sign-problem free {\em ab initio} Monte Carlo simulations of hypernuclear system containing arbitrary number of neutrons, protons and $\Lambda$ hyperons, consistently including all relevant two- and three-body forces. 
This overcomes the need for the symmetry-energy approximation, which has long been used to interpolate between symmetric nuclear matter and neutron matter, thereby enabling a direct and consistent treatment of $\beta$-stable hypernuclear matter without using external tools such as Gaussian process emulators~\cite{Keller:2022crb,Keller:2024snt}. 
Our results show that neutron stars with masses around $2M_\odot$ can accommodate substantial hyperon fractions — up to nearly 39\% in the stellar core — in contrast to earlier studies~\cite{Lonardoni:2014bwa} that ruled out hyperons to satisfy the mass constraint. 
The agreement with hypernuclei experimental data and various recent astrophysical observations demonstrates that the long-standing {\em hyperon puzzle} can be resolved within our state-of-the-art {\em ab initio} approach. 
This establishes a promising path toward a comprehensive microscopic understanding of strangeness in nuclear systems and its essential role in shaping the structure and observable properties of neutron stars.

\medskip

%\clearpage

{\it We are grateful for discussions with members and partners of the Nuclear Lattice Effective Field Theory Collaboration. We thank Liewen Chen, Tetsuo Hatsuda, Baoan Li, Haozhao Liang, Jie Meng and Sibo Wang for helpful discussions. We acknowledge funding by the European Research Council (ERC) under the European Union's
Horizon 2020 research and innovation programme (AdG EXOTIC, grant agreement No. 101018170) and by the MKW
NRW under the funding code NW21-024-A.
The work of SE was further supported by  the Scientific and Technological Research Council of Turkey (TUBITAK project no. 123F464).
The work of UGM was further supported by CAS through the President's International Fellowship Initiative (PIFI) (Grant No. 2025PD0022).
The authors gratefully acknowledge the Gauss Centre for Supercomputing e.V. (www.gauss-centre.eu)
for funding this project by providing computing time on the GCS Supercomputer JUWELS
at J\"ulich Supercomputing Centre (JSC). Furthermore, the authors gratefully acknowledge the computing time provided on the high-performance computer
HoreKa by the National High-Performance Computing Center at KIT (NHR@KIT). This center is
jointly supported by the Federal Ministry of Education and Research and the Ministry of Science,
Research and the Arts of Baden-Württemberg, as part of the National High-Performance Computing
(NHR) joint funding program (https://www.nhr-verein.de/en/our-partners). HoreKa is partly funded
by the German Research Foundation (DFG).
}

%% file: 3-method.tex
\section{Nuclear Lattice Effective Field Theory}

%\subsection{Lattice Formalism and Auxiliary Field Formulation for Hypernuclear Systems}
%---------------------------------------------------}
\label{sec:Lattice-Hamiltonian}

Nuclear Lattice effective field theory (NLEFT) is a quantum many-body approach that combines the framework of effective field theory (EFT) with powerful Monte Carlo techniques~\cite{Lee:2008fa,Lahde:2019npb}. 
It has been successfully applied in pionless EFT at leading order (LO) to describe atomic nuclei~\cite{Borasoy:2005yc} and neutron matter~\cite{Lee:2004qd}, and in chiral EFT at next-to-next-to-leading order (N2LO) to achieve the first {\em ab initio} calculations of the Hoyle state in $^{12}$C~\cite{Epelbaum:2011md} and $\alpha$–$\alpha$ scattering~\cite{Elhatisari:2015iga}. 
More recently, NLEFT has been extended to compute nuclear properties and the equation of state (EoS) of neutron and symmetric nuclear matter at next-to-next-to-next-to-leading order (N3LO)~\cite{Elhatisari:2022zrb}. 
A particularly efficient formulation is based on Wigner’s SU(4) spin–isospin symmetry ~\cite{Wigner:1936dx} with only four parameters, which reproduces bulk properties of light- and medium-mass nuclei and the neutron matter EoS to within a few percent~\cite{Lu:2018bat}. 
This formulation has enabled applications ranging from the first {\em ab initio} thermodynamics of nuclear clustering~\cite{Lu:2019nbg} and studies of clusters in hot dilute matter~\cite{Ren:2023ued} to the emergent geometry of $^{12}$C~\cite{Shen:2021kqr,Shen:2022bak} and the resolution of the $\alpha$-particle monopole transition form factor puzzle~\cite{Meissner:2023cvo}. 
Building on the success of EFT with Wigner’s SU(4) spin–isospin symmetry -- referred to here as the minimal nuclear interaction -- we employ pionless EFT at LO for nucleons~\cite{Konig:2016utl} as the basis of our calculations. 
For the hyperon-nucleon and hyperon-hyperon interactions, we also utilize minimal interactions. 
This provides an efficient framework for determining the properties of hypernuclei and hypernuclear matter~(HNM). 

The Hamiltonian in this work is defined as,
\begin{align}
H= & H_{\rm free}+\frac{c_{NN}}{2}\sum_{\vec{n}}
\,:\,\left[
\tilde{\rho}(\vec{n})
\right]^2
\,:\,
+\frac{c_{NN}^{T}}{2}\sum_{I,\vec{n}}
\,:\,
\left[
\tilde{\rho}_{I}(\vec{n})
\right]^2 
\,:\,
\nonumber\\
& + c_{N\Lambda}\sum_{\vec{n}}
\,:\,
\tilde{\rho}(\vec{n})
\tilde{\xi}(\vec{n})
\,:\,
+ \frac{c_{\Lambda\Lambda}}{2}\sum_{\vec{n}}
\,:\,
\left[
\tilde{\xi}(\vec{n})
\right]^2
\,:\,
\nonumber\\
 &
+V^{\rm GIR}_{NN} 
+V^{\rm GIR}_{N\Lambda} 
+V^{\rm GIR}_{\Lambda\Lambda} 
+V_{\rm Coulomb}
\nonumber\\
 &
+V_{NNN} 
+V_{NN\Lambda} 
+V_{N\Lambda\Lambda}
\,,
\label{eq:H-001}
\end{align}
where $H_{\rm free}$ is the kinetic energy term for nucleons and $\Lambda$ hyperons, the $::$ symbol indicates normal ordering, $c_{NN}$ is the coupling constant of the short-range two-nucleon interaction, $c_{NN}^{T}$ is the coupling constant of the isospin-dependent short-range two-nucleon interaction (that breaks SU(4) symmetry but is required to describe nuclear matter), $c_{N\Lambda}$ ($c_{\Lambda\Lambda}$) is the coupling constant of the short-ranged hyperon-nucleon (hyperon-hyperon) interaction, and $\tilde{\rho}$ ($\tilde{\xi}$) is nucleon (hyperon) density operator, that is smeared both locally and non-locally,
\begin{align}
\tilde{\rho}(\vec{n}) = \sum_{i,j=0,1} 
\tilde{a}^{\dagger}_{i,j}(\vec{n}) \, \tilde{a}^{\,}_{i,j}(\vec{n})
+
s_{\rm L}
 \sum_{|\vec{n}-\vec{n}^{\prime}|^2 = 1}
 \,
 \sum_{i,j=0,1} 
\tilde{a}^{\dagger}_{i,j}(\vec{n}^{\prime}) \, \tilde{a}^{\,}_{i,j}(\vec{n}^{\prime})
\,,
\end{align}
\begin{align}
\tilde{\rho}_{I}(\vec{n})= \sum_{i,j,j^{\prime}=0,1} 
\tilde{a}^{\dagger}_{i,j}(\vec{n}) \,\left[\tau_{I}\right]_{j,j^{\prime}} \, \tilde{a}^{\,}_{i,j^{\prime}}(\vec{n})
+
s_{\rm L}
 \sum_{|\vec{n}-\vec{n}^{\prime}|^2 = 1}
 \,
  \sum_{i,j,j^{\prime}=0,1} 
\tilde{a}^{\dagger}_{i,j}(\vec{n}^{\prime}) \,\left[\tau_{I}\right]_{j,j^{\prime}} \, \tilde{a}^{\,}_{i,j^{\prime}}(\vec{n}^{\prime})\,,
\end{align}
\begin{align}
\tilde{\xi}(\vec{n}) = \sum_{i=0,1} 
\tilde{b}^{\dagger}_{i}(\vec{n}) \, \tilde{b}^{\,}_{i}(\vec{n})
+
s_{\rm L}
 \sum_{|\vec{n}-\vec{n}^{\prime}|^2 = 1}
 \,
 \sum_{i=0,1} 
\tilde{b}^{\dagger}_{i}(\vec{n}^{\prime}) \, \tilde{b}^{\,}_{i}(\vec{n}^{\prime})
\,.
\end{align}
The smeared annihilation and creation operators, $\tilde{a}$ ($\tilde{b}$) and $\tilde{a}^{\dagger}$ ($\tilde{b}^{\dagger}$) for nucleons (hyperons),  with spin $i = 0, 1$ (up, down) and isospin $j = 0, 1$ (proton, neutron) indices, are given by
\begin{align}
\tilde{a}_{i,j}(\vec{n})=a_{i,j}(\vec{n})+s_{\rm NL}\sum_{|\vec{n}^{\prime}-\vec{n}|=1}a_{i,j}(\vec{n}^{\prime})\,,
\label{eqn:rho-local-nonlocal}
\end{align}
\begin{align}
\tilde{b}_{i}(\vec{n})=b_{i}(\vec{n})+s_{\rm NL}\sum_{|\vec{n}^{\prime}-\vec{n}|=1}b_{i}(\vec{n}^{\prime}).
\label{eqn:xi-local-nonlocal}
\end{align}
$V_{\text{Coulomb}}$ represents the Coulomb interaction in Eq.~(\ref{eq:H-001}). 
The nonlocal smearing applied on the lattice induces a center-of-mass momentum dependence that breaks Galilean invariance. To restore it, we include $V^{\text{GIR}}_{NN}$, $V^{\text{GIR}}_{N\Lambda}$, and $V^{\text{GIR}}_{\Lambda\Lambda}$ in Eq.~(\ref{eq:H-001}), corresponding to Galilean invariance restoration~(GIR) interactions. Note that the smearing effectively introduces
higher-order (derivative) interactions.

Recent \emph{ab-initio} nuclear structure and scattering calculations have revealed the significant impact of locally smeared interactions on nuclear binding~\cite{Elhatisari:2016owd}. Hence, the three-baryon interactions $V_{NNN}$, $V_{NN\Lambda}$, and $V_{N\Lambda\Lambda}$ utilized in Eq.~(\ref{eq:H-001}) are defined with two different choices of local smearing,
\begin{align}\label{eq:VNNN}
V_{NNN}
= 
\frac{c_{NNN}^{(d_1)}}{6} 
\,
\sum_{\vec{n}} 
\,:\,
\left[
\rho^{(d_1)}(\vec{n})
\right]^3
\,:\,
+
\frac{c_{NNN}^{(d_2)}}{6} 
\,
\sum_{\vec{n}} 
\,:\,
\left[
\rho^{(d_2)}(\vec{n})
\right]^3
\,:\,
\,,
\end{align}
where the parameter $d_i$ denotes the range of the local smearing with $0 \leq d_1 < d_2 \leq 3$ (in lattice units). Similarly, the three-baryon interactions $V_{NN\Lambda}$ are defined as,
\begin{align}
V_{NN\Lambda}
=
\sum_{i = 1,2}
\frac{c_{NN\Lambda}^{(d_i)}}{2}
\,
\sum_{\vec{n}}
\,:\,
\left[
\rho^{(d_i)}(\vec{n})
\right]^2 \xi^{(d_i)}(\vec{n})
\,:\,,
\label{eqn:V-NNY}
\end{align}
and $V_{N\Lambda\Lambda}$,
\begin{align}
V_{N\Lambda\Lambda}
=
\sum_{i = 1,2}
\frac{c_{N\Lambda\Lambda}^{(d_i)}}{2}
\,
\sum_{\vec{n}} \,
 \,:\,
 \rho^{(d_i)}(\vec{n})  \,
\left[
\xi^{(d_i)}(\vec{n})
\right]^2
\,:\,,
\label{eqn:V-NYY}
\end{align}
where ${\rho}$ (${\xi}$) is then a purely locally smeared nucleon (hyperon) density operator with annihilation and creation operators, ${a}$ (${b}$) and ${a}^{\dagger}$ (${b}^{\dagger}$) for nucleons (hyperons),
\begin{align}
{\rho}^{(d)}(\vec{n}) = \sum_{i,j=0,1} 
{a}^{\dagger}_{i,j}(\vec{n}) \, {a}^{\,}_{i,j}(\vec{n})
+
s^{\rm 3B}_{\rm L}
 \sum_{|\vec{n}-\vec{n}^{\prime}|^2 = 1}^{d}
 \,
 \sum_{i,j=0,1} 
{a}^{\dagger}_{i,j}(\vec{n}^{\prime}) \, {a}^{\,}_{i,j}(\vec{n}^{\prime})
\,,
\label{eqn:rho-local}
\end{align}
\begin{align}
{\xi}^{(d)}(\vec{n}) = \sum_{i=0,1} 
{b}^{\dagger}_{i}(\vec{n}) \, {b}^{\,}_{i}(\vec{n})
+
s^{\rm 3B}_{\rm L}
 \sum_{|\vec{n}-\vec{n}^{\prime}|^2 = 1}^{d}
 \,
 \sum_{i=0,1} 
{b}^{\dagger}_{i}(\vec{n}^{\prime}) \, {b}^{\,}_{i}(\vec{n}^{\prime})
\,.
\label{eqn:xi-local}
\end{align}
Here, $s^{\rm 3B}_{\rm L}$ defines the strength of the local smearing. In our analysis of locally smeared three-baryon forces, we exclusively consider smearing with ranges $d \leq 3$ in lattice units. 
In this work, the $NN$ and $NNN$ interactions are fitted to the experimental scattering phase shifts and the saturation properties of symmetric nuclear matter by considering all possible combinations of $d_1$ and $d_2$ with $0 \leq d_1 < d_2 \leq 3$. 
For the $N\Lambda$ and $\Lambda\Lambda$ interaction, we fit experimental cross-section data and phase shift derived from chiral EFT at next-to-leading order~\cite{Haidenbauer:2015zqb}. 
The numerical values of the various LECs for these interactions and lattice parameters are given in Refs.~\cite{Tong:2024jvs,Tong:2025sui}. 
To constrain the LECs of the $NN\Lambda$ and $N\Lambda\Lambda$ three-baryon forces in the present work, the separation energies of light hypernuclei $_\Lambda^5$He, $_\Lambda^8$Be, $_\Lambda^9$Be, $_{\Lambda\Lambda}^6$He, and $_{\Lambda\Lambda}^{10}$Be are used,
\begin{equation}
  \begin{split}
    &B_\Lambda(_\Lambda^A Z)=E(^{A-1}Z)-E(_\Lambda^{A}Z),\\
    &B_{\Lambda\Lambda}(_{\Lambda\Lambda}^A Z)=E(^{A-2}Z)-E(_{\Lambda\Lambda}^{A}Z),
  \end{split}
\end{equation}
where $E$ is the energy of the system, $A$ its atomic number and $Z$ its charge.
Through this process, we arrive at six different sets of interactions, enabling us to quantify the theoretical uncertainty of our calculations. 
%In our calculations, we use a spatial lattice spacing of $a=1.1$ fm, temporal spacing $a_t=0.2$ fm, and smearing parameters $s_{\rm L}=0.06$ and $s_{\rm NL}=0.6$ for two-baryon interactions, with $s^{\rm 3B}_{\rm L}=0.06$ for three-baryon forces. For HNM we employ $6.6$ fm lattices with average twisted boundary conditions (ATBC) to suppress finite-volume effects~\cite{Li:2019ldq}.

In the present work, we use a newly developed approach that allows for the efficient investigation of hypernuclear systems with an arbitrary number of neutrons, protons, and hyperons. 
We employ the auxiliary-field quantum Monte Carlo (AFQMC) method in our lattice simulations, which strongly suppresses sign oscillations (see Ref.~\cite{Lahde:2019npb} for a review). 
AFQMC reformulates the $A$-body Schrödinger equation through a Hubbard–Stratonovich transformation that introduces auxiliary fields to decouple particle densities, enabling efficient Monte Carlo sampling. 
The discrete auxiliary field formulation for the SU(4) symmetric short ranged two-nucleon interaction given in Eq.~(\ref{eq:H-001}),
\begin{align}
   : \exp \left( -\frac{a_{t} \, c_{NN}}{2} \, \tilde{\rho}^2  
          \right) :
=
\sum_{k = 1}^{3} \, w_{k} \, 
: \exp \left( \sqrt{-a_{t} \, c_{NN}}  \, s_{k} \, \tilde{\rho} \right) \, :
\label{eqn:AFQMC-NN}
\end{align}
where $a_{t}$ is the temporal lattice spacing. From a Taylor expansion of Eq.~(\ref{eqn:AFQMC-NN}) we determine the constants $s_{k}$ and $w_k$ as $s_{1} = -s_{3}=\sqrt{3}$, $s_{2} = 0$, $w_{1} = w_3 = 1/6$ and $w_2 = 2/3$.
%The nucleon-nucleon interaction given in Eq.~(\ref{eqn:AFQMC-NN}) obeys the Wigner SU(4) symmetry~\cite{Wigner:1936dx}, which arises from the realization that the combined spin ($S$) and isospin ($T$) degrees of freedom of nucleons can be described by a single unified symmetry group. Since we use minimal forces for the hyperon-nucleon and hyperon-hyperon interactions, we now aim to derive an auxiliary field formulation for systems including neutrons, protons and $\Lambda$ hyperons. This derivation involves replacing the isospin SU$_T(2)$ with flavor SU$_F(3)$ within Wigner's SU(4) symmetry framework, and the combined spin ($S$) and flavor ($F$) invariance ultimately leads to the SU(6) symmetry~\cite{Gursey:1964htz}. However, the fact that the strengths of the nucleon-nucleon and hyperon-nuclear interactions are different is breaking this SU(6) symmetry, and there is no longer an approximate symmetry similar to Wigner’s SU(4) symmetry used in Eq.~(\ref{eqn:AFQMC-NN}). Nevertheless, in the following we exploit the fact that $|c_{NN}|>|c_{N\Lambda}|>§|c_{\Lambda\Lambda}|$, which is allowing us to introduce an auxiliary field formulation with an approximate SU(6) symmetry that protects our simulations including $\Lambda$ hyperons against strong sign oscillations.
The spin and isospin independent two-baryon interactions in Eq.~(\ref{eq:H-001}) is expressed as,
\begin{align}
V_{\rm 2B} = \frac{c_{NN}}{2}\sum_{\vec{n}}
\,:\,
\left[
\tilde{\rho}(\vec{n})
\right]^2
\,:\,
+ c_{N\Lambda}\sum_{\vec{n}}
\,:\, 
\tilde{\rho}(\vec{n})
\tilde{\xi}(\vec{n})
\,:\,
+ \frac{c_{\Lambda\Lambda}}{2}\sum_{\vec{n}}
\,:\,
\left[
\tilde{\xi}(\vec{n})
\right]^2
\,:\,
\,,
\label{eqn:NY-Potential-000}
\end{align}
and this potential (\ref{eqn:NY-Potential-000}) can be rewritten in the following form,
\begin{align}
V_{\rm 2B} = \frac{c_{NN}}{2}\sum_{\vec{n}}
\,:\,
\left[
\tilde{\slashed{\rho}}(\vec{n})
\right]^2
\,:\,
+
\frac{1}{2}
\left(
c_{\Lambda\Lambda}
-\frac{c_{N\Lambda}^2}{c_{NN}} 
\right)
\sum_{\vec{n}}
\,:\,
\left[
\tilde{\xi}(\vec{n})
\right]^2 
\,:\,
\,,
\label{eqn:NY-Potential-010}
\end{align}
where $\tilde{\slashed{\rho}}$ is defined as,
\begin{align}
\tilde{\slashed{\rho}} = 
\tilde{\rho} + 
\frac{c_{ N\Lambda}}{c_{NN}} \,  \tilde{\xi}
\,.
\label{eqn:Density-rho-bar}
\end{align}
%In Eq.~(\ref{eqn:NY-Potential-010}) the leading contribution comes from the first term in the right-hand side and it is treated non-perturbatively, while the remaining term is computed using first-order perturbation theory. Hence, 
We define a new Hubbard-Stratonovich transformation for the Eq.~(\ref{eqn:NY-Potential-010}), enabling the simulations of systems consisting of an arbitrary number of nucleons and arbitrary number of $\Lambda$ hyperons with a single auxiliary field,
\begin{align}
   : \exp \left( -\frac{a_{t} \, c_{NN}}{2} \, \tilde{\slashed{\rho}}^2\right) :
=
\sum_{k = 1}^{3} \, w_{k} \, 
: \exp \left( \sqrt{-a_{t} \, c_{NN}}  \, s_{k} \,  \tilde{\slashed{\rho}} \right) \,. :
\label{eqn:AFQMC-NY}
\end{align}
The solution for the auxiliary field variables $s_{k}$ and weights $w_{k}$ is consistent with systems containing only nucleons. Such a transformation was recently used to achieve the first sign-problem free implementation of spin–orbit coupling for shell evolution~\cite{Niu:2025uxk}.

\section{Equation of State and neutron star properties}

In this section, we start by giving the expressions used to describe the EoS and neutron star properties, focusing on hypernuclear matter composed of neutrons, protons, and $\Lambda$ hyperons. 
The particle fractions of neutrons, protons, and $\Lambda$ hyperons are defined as $x_n=\rho_n/\rho$, $x_p=\rho_p/\rho$, and $x_\Lambda=\rho_\Lambda/\rho$, where $\rho=\rho_n+\rho_p+\rho_\Lambda$ represents the total baryon density. 
The energy per particle for $\beta$-stable HNM can be expressed as
\begin{equation}
  e_{\rm HNM}(x_n,x_p,x_\Lambda,x_e,x_\mu) = \frac{E_{\rm HNM}(x_n,x_p,x_\Lambda)}{N_{\rm tot}} + x_n m_n + x_p m_p + x_\Lambda m_\Lambda + e_e + e_\mu,
\end{equation}
where $E_{\rm HNM}$ is the total energy, $N_{\rm tot}=N_n+N_p+N_\Lambda$ the total number of baryons, and $e_e$, $e_\mu$ the contributions from electrons and muons. 
To evaluate $e_{\scriptscriptstyle \rm HNM}$, the total energy $E_{\rm HNM}$ must be determined for arbitrary numbers of neutrons, protons, and $\Lambda$ hyperons -- a task made feasible by our NLEFT, which enables sign-problem free simulations of such mixed systems. 
This capability is essential for constructing the $\beta$-stable EoS of neutron star matter.

\begin{figure}[htbp]
  \centering
  \includegraphics[height=9.5 cm,valign=t]{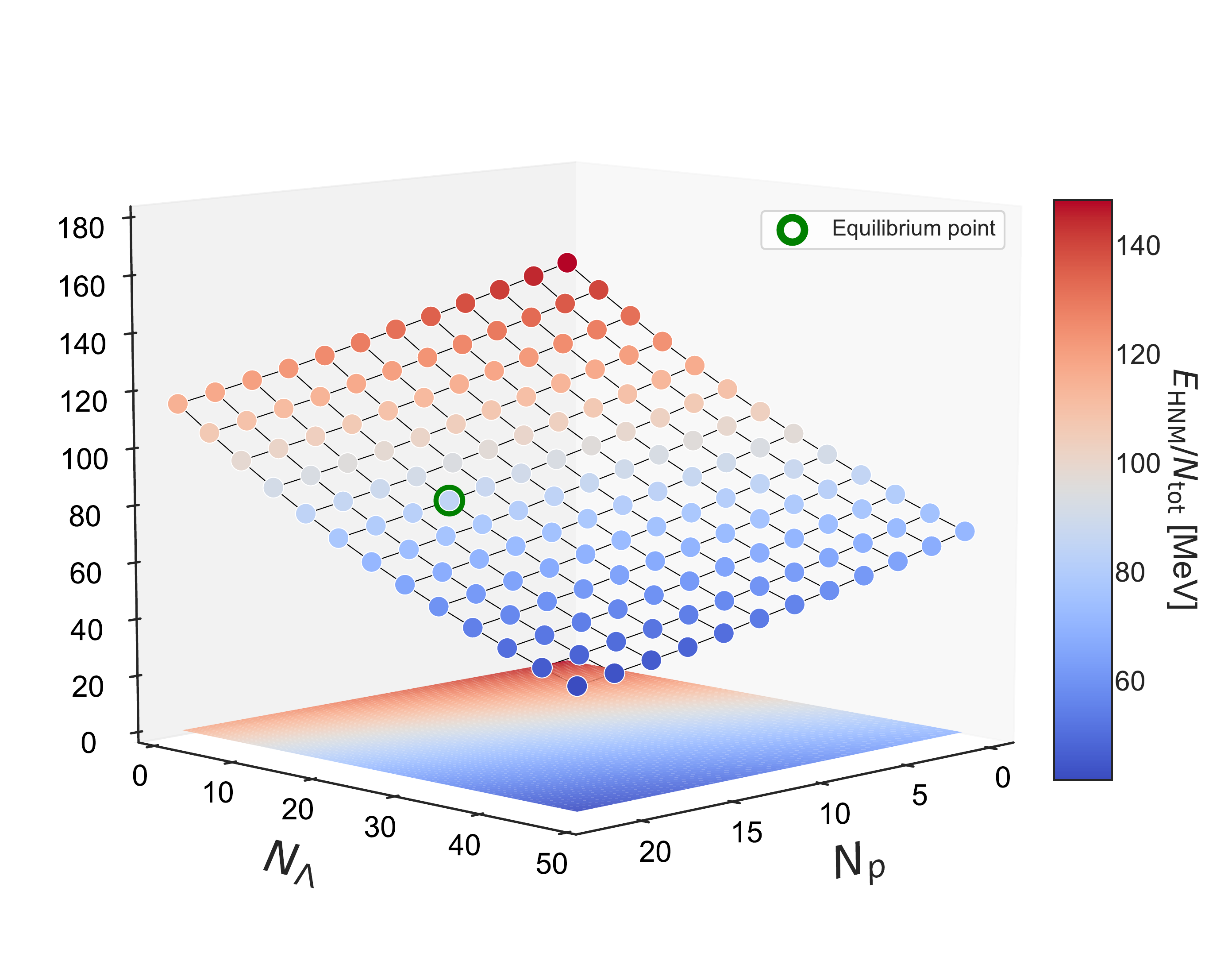}
  \caption{{\bf 3D plot showing the energy for HNM.} Energy per baryon $E_{\rm HNM}/N_{\rm tot}$ as a function of proton number $N_p$ and $\Lambda$ hyperon number $N_\Lambda$ at fixed density $\rho=0.6$ fm$^{-3}$. The green circle indicates the equilibrium composition of $\beta$-stable matter at this density. The color scale denotes the magnitude of the energy per baryon.}
  \label{figS1}
\end{figure}

Figure~\ref{figS1} displays the energy per baryon of HNM as a function of the proton number $N_p$ and the $\Lambda$ hyperon number $N_\Lambda$ at fixed density $\rho=0.6$~fm$^{-3}$. 
One can observe that the energy decreases as either $N_p$ or $N_\Lambda$ increases, highlighting the crucial role played by protons and hyperons in reducing the total energy of the system. 
This demonstrates quantitatively that the inclusion of non-nucleonic degrees of freedom significantly modifies the energetics of dense matter and hence the composition of $\beta$-stable matter at high density. 
This figure also illustrates the methodological advances of our work. 
For the first time, within the NLEFT framework we have carried out {\em ab initio} Monte Carlo simulations of a fully mixed system containing an arbitrary number of neutrons, protons, and $\Lambda$ hyperons. 
Such simulations with multiple fermionic degrees of freedom are extremely demanding computationally, but our new AFQMC formulation makes them feasible while avoiding the parameterized interpolations employed in earlier AFDMC studies~\cite{Lonardoni:2014bwa}. 
By eliminating these approximations, our approach provides a controlled microscopic description of HNM and establishes the foundation for constructing the $\beta$-stable EoS.

The energy density $\varepsilon_{\scriptscriptstyle \rm HNM}$ is defined as $\varepsilon_{\scriptscriptstyle \rm HNM} = \rho e_{\scriptscriptstyle \rm HNM}$. 
The chemical potentials of the neutrons, the protons and the $\Lambda$s, $\mu_n$, $\mu_p$, and $\mu_\Lambda$, are obtained from the thermodynamic relations
\begin{equation}
  \mu_n(\rho,x_p,x_\Lambda)=\frac{\partial \varepsilon_{\scriptscriptstyle \rm HNM}}{\partial \rho_n}, ~~\mu_p(\rho,x_p,x_\Lambda)=\frac{\partial \varepsilon_{\scriptscriptstyle \rm HNM}}{\partial \rho_p}, ~~\mu_\Lambda(\rho,x_p,x_\Lambda)=\frac{\partial \varepsilon_{\scriptscriptstyle \rm HNM}}{\partial \rho_\Lambda},
\end{equation}
the particle fractions, $x_i(\rho)$, with $i=n,p,\Lambda$, are determined by imposing the chemical-equilibrium condition in Eq.~\eqref{equ-chemicalequilibrium} together with charge neutrality. 
This procedure defines the threshold density $\rho_\Lambda^{\rm th}$, marking the point where $x_\Lambda(\rho)$ first departs from zero. 
The open green circle in Fig.~\ref{figS1} marks the equilibrium configuration of HNM at $\rho=0.6$ fm$^{-3}$. 
Extracting such an equilibrium point is highly nontrivial in {\em ab initio} Monte Carlo simulations: it requires exploring the full three-dimensional space of neutrons, protons, and $\Lambda$ hyperons, a task that involves extremely demanding computations. 

\begin{figure}[htbp]
  \centering
  \includegraphics[height=10 cm,valign=t]{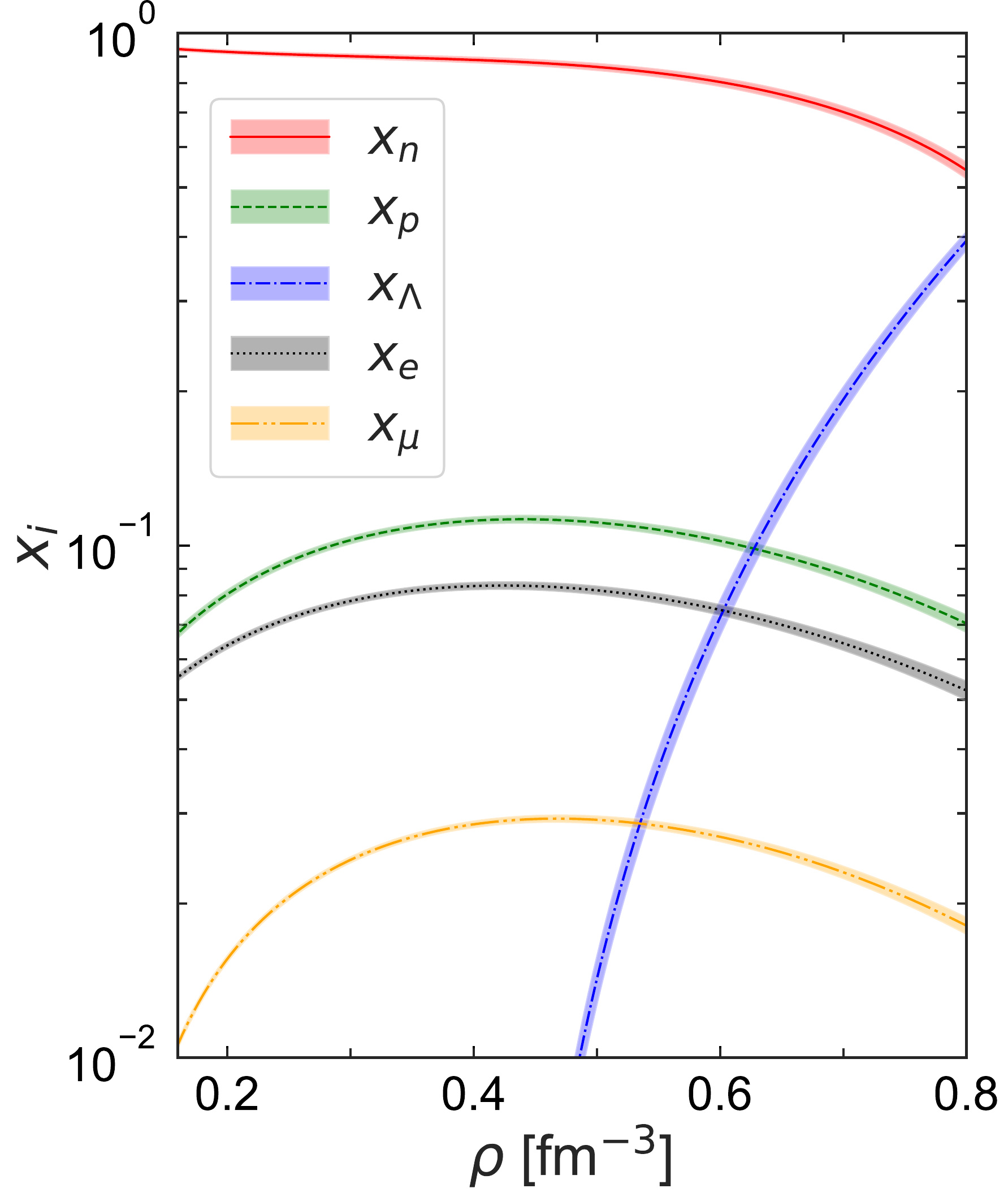}
  \caption{{\bf Particle fractions in $\beta$-stable HNM.} Density dependence of the particle fractions $x_i(\rho)$ for neutrons ($x_n$), protons ($x_p$), $\Lambda$ hyperons ($x_\Lambda$), electrons ($x_e$), and muons ($x_\mu$) are shown. Bands indicate statistical and systematic uncertainties of the NLEFT calculations.
  }
  \label{figS2}
\end{figure}

Figure~\ref{figS2} shows the density dependence of the particle fractions $x_i(\rho)$ in $\beta$-stable HNM. 
At low densities the composition is dominated by neutrons, with protons, electrons, and muons providing the necessary charge neutrality. 
With increasing density, the proton fraction rises to values above $\sim 10\%$ before gradually decreasing at higher densities. 
A similar trend is seen for the leptons: both electrons and muons increase initially as density grows, but their fractions are eventually reduced.  
These trends are in full qualitative agreement with earlier microscopic studies based on non-relativistic Brueckner–Hartree–Fock theory~\cite{Logoteta:2019utx,Vidana:2024ngv}. 
Another key feature is the onset of $\Lambda$ hyperons at the threshold density $\rho_\Lambda^{\rm th}\simeq0.490(0.002)(0.005)$ fm$^{-3}$, where $x_\Lambda(\rho)$ first deviates from zero. 
Beyond this point, the $\Lambda$ fraction increases rapidly with density, while the neutron fraction decreases, reflecting that the conversion of neutrons into hyperons becomes energetically favored once chemical equilibrium is established. 
Above five times nuclear saturation density, the predicted $\Lambda$ fraction reaches nearly 39\%, demonstrating that hyperons constitute a major component of dense matter in neutron star cores.

Finally, the pressure $P(\rho)$ of HNM is obtained from the energy density,
\begin{equation}\label{ns-eos}
    P(\rho)=\rho^2\frac{d}{d\rho}\frac{\varepsilon_{\scriptscriptstyle \rm HNM}}{\rho}.%=\sum_{i=n,p,\Lambda}\rho_i\mu_i-\varepsilon_{\scriptscriptstyle \rm HNM}.
\end{equation}
Once the EoS of HNM in the form $P(\varepsilon)$ is obtained in Eq.~\eqref{ns-eos}, it can be used to derive additional properties of dense matter. 
In particular, the squared speed of sound, $c_s^2 = \mathrm{d}P/\mathrm{d}\varepsilon$, characterizes the stiffness of the EoS and directly impacts the neutron star structure. 
Another useful quantity is the normalized trace anomaly, $\Delta=1/3-P/\varepsilon$, a dimensionless measure of conformality that affects the sound velocity via both derivative and non-derivative terms~\cite{Fujimoto:2022ohj,Annala:2023cwx,Cai:2024oom}. 
It quantifies the deviation from conformal behavior and serves as a sensitive probe of the underlying microscopic degrees of freedom at supra-nuclear densities.

\begin{figure}[htbp]
  \centering
  \includegraphics[height=7 cm,valign=t]{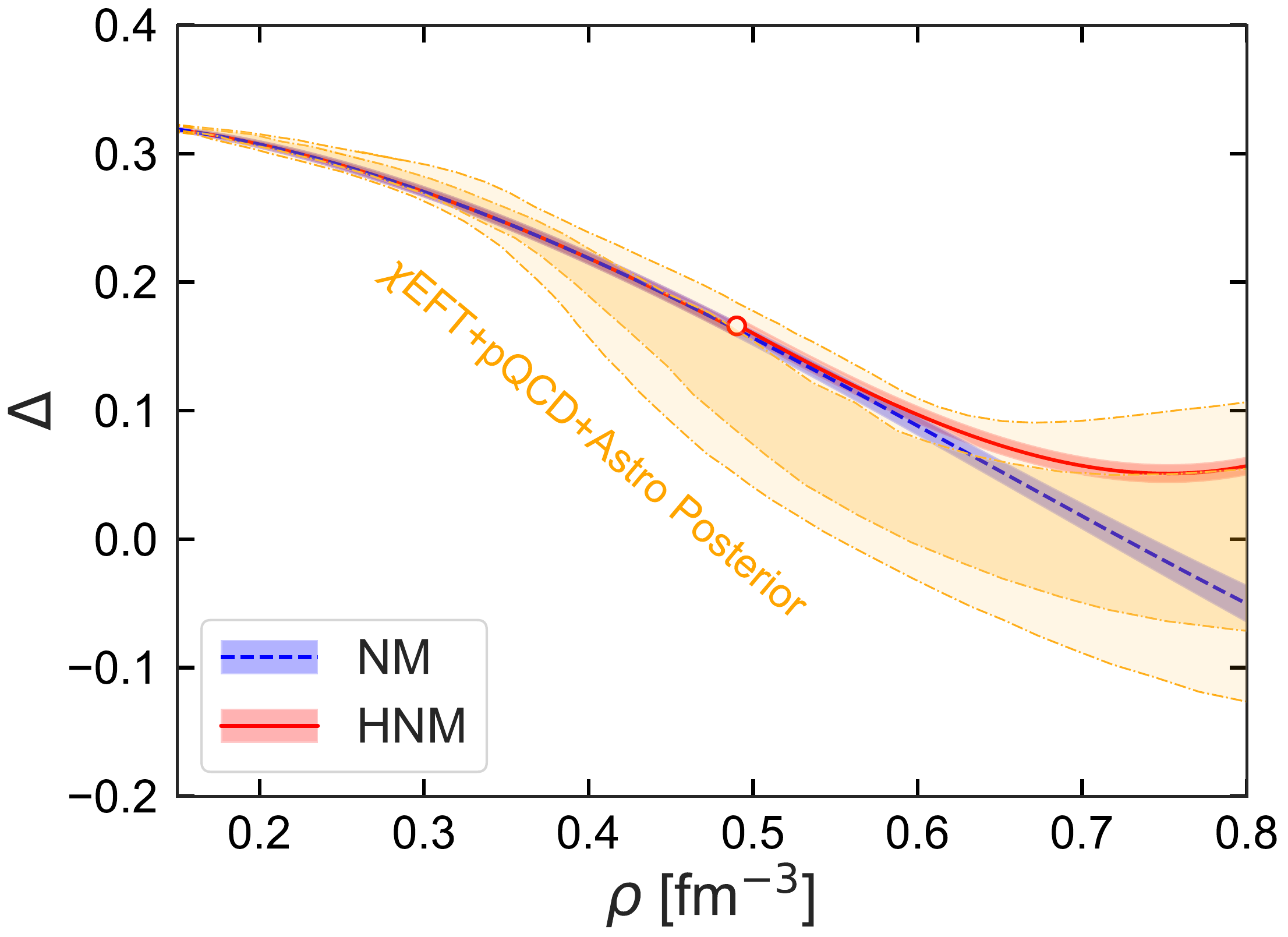}
  \caption{{\bf Normalized trace anomaly.} The normalized trace anomaly $\Delta$ as a function of density $\rho$. The legend is the same as of Fig.~\ref{fig2}. The dark and light orange bands represent the 68\% and 95\% credible intervals, respectively, obtained from a four-segment speed-of-sound interpolation constrained by Bayesian inference combining theoretical calculations and astrophysical observations~\cite{Annala:2023cwx,Hebeler:2013nza,Drischler:2020hwi,Kurkela:2009gj,Gorda:2021znl}. 
  }
  \label{figS3}
\end{figure}

Figure~\ref{figS3} shows the normalized trace anomaly $\Delta$ as a function of density $\rho$ for both NM and HNM. 
At low densities, both NM and HNM display positive $\Delta$, reflecting significant deviations from conformality. 
As the density increases, $\Delta$ decreases and approaches zero, indicating a gradual trend toward conformal behavior at supra-nuclear densities. 
The inclusion of $\Lambda$ hyperons shifts the trajectory of $\Delta$ relative to the purely nucleonic case, with HNM exhibiting a systematically softer behavior at high densities. 
This reflects the additional degrees of freedom carried by hyperons, which lower the pressure at fixed energy density. 
The predicted curves for both NM and HNM remain well within the 68\% and 95\% credible intervals inferred from Bayesian analyses constrained by $\chi$EFT at low densities, perturbative QCD (pQCD) at asymptotically high densities, and astrophysical observations~\cite{Annala:2023cwx,Hebeler:2013nza,Drischler:2020hwi,Kurkela:2009gj,Gorda:2021znl}. 
This agreement demonstrates that our {\em ab initio} results not only capture the essential features of neutron star matter across a wide density range, but also highlight the quantitative role of hyperons in driving the system toward conformality in neutron star cores.

The neutron star mass and radius can be described by the Tolman-Oppenheimer-Volkoff (TOV) equations~\cite{Tolman:1939jz,Oppenheimer:1939ne}
\begin{subequations}\label{ns-toveq}
  \begin{align}
    \frac{dP(r)}{dr}=&\ -\frac{[P(r)+\varepsilon(r)][M(r)+4\pi r^3P(r)]}{r[r-2M(r)]}, \\
	\frac{dM(r)}{dr} =&\ 4\pi r^2\varepsilon(r),
  \end{align}
\end{subequations}
where $P(r)$ is the pressure at radius $r$ and $M(r)$ is the total mass inside a sphere of radius $r$. 
Furthermore, to solve the TOV equations, the EoS must cover the entire structure of the neutron star, from the crust to the core. In this work, we mainly focus on discussing the core region within the NLEFT. For the crust, we adopt the well-established EoSs formulated by Baym, Pethick, Sutherland (BPS)~\cite{Baym1971-BPS} and by Baym, Bethe, and Pethick (BBP)~\cite{Baym1971-BBP}.
Another important property of a neutron star, the tidal deformability $\Lambda$, is defined as
\begin{equation}
  \Lambda = \frac{2}{3}k_2 C^{-5},
\end{equation}
which represents the mass quadrupole moment response of a neutron star to the strong gravitational field induced by its companion. 
$C=M/R$ is the compactness parameter, $M$ and $R$ are the neutron star mass and radius, and $k_2$~is the second love number.